# Code Spectrum and Reliability Function: Binary Symmetric Channel

Marat V. Burnashev

Institute for Information Transmission Problems,

Russian Academy of Sciences, Moscow, Russia

Email: burn@iitp.ru

### Abstract

A new approach for upper bounding the channel reliability function using the code spectrum is described. It allows to treat in a unified way both a low and a high rate cases. In particular, the earlier known upper bounds are improved, and a new derivation of the sphere-packing bound is presented.

## I. Introduction and main results

We consider a binary symmetric channel (BSC) with crossover probability $0 < p < 1/2$ and $q = 1 - p$. Let $F^n$ denote the set of all $2^n$ binary $n$-tuples, and, $d(\boldsymbol{x}, \boldsymbol{y})$, $\boldsymbol{x}, \boldsymbol{y} \in F^n$, denote the Hamming distance between $\boldsymbol{x}$ and $\boldsymbol{y}$. A subset $\mathcal{C} = \{\boldsymbol{x}_1, \ldots, \boldsymbol{x}_M\} \subseteq F^n$, $M = 2^{Rn}$ is called a $(R, n)$–code of the rate $R$. The *minimum distance* of the code $\mathcal{C}$ is $d(\mathcal{C}) = \min\{d(\boldsymbol{x}_i, \boldsymbol{x}_j) : i \neq j\}$. Everywhere below $\log z = \log_2 z$.

Cardinality of a set $A$ is denoted by $|A|$. The *distance distribution* $B(\mathcal{C}) = (B_0, B_1, \ldots, B_n)$ of the code $\mathcal{C}$ is the $(n+1)$–tuple with components

$$B_i = |\mathcal{C}|^{-1} \left|\{(\boldsymbol{x}_j, \boldsymbol{x}_k) : d(\boldsymbol{x}_j, \boldsymbol{x}_k) = i\}\right|, \qquad i = 0, 1, \ldots, n. \tag{1}$$

Let $h_2(x) = -x \log_2 x - (1-x) \log_2(1-x)$ be the binary entropy function, and $h_2^{-1}(y)$, $y \in [0, 1/2]$, be its inverse function.

The BSC *reliability function* $E(R, p)$ is defined as follows [1, 2, 3]

$$E(R, p) = \limsup_{n \to \infty} \frac{1}{n} \ln \frac{1}{P_e(R, n, p)},$$

where $P_e(R, n, p)$ is the minimal possible decoding error probability $P_e$ for $(R, n)$–code. After the fundamental results of the paper [1], a further improvements of a various bounds for $E(R, p)$ have been obtained in [2 – 8]. Generally, on the exact form of the function $E(R, p)$ it is known only that [1, 2, 3]

$$E(0, p) = \frac{1}{4} \ln \frac{1}{4pq}, \qquad E(R, p) = E_{\mathrm{sp}}(R, p), \quad R_{\mathrm{crit}}(p) \leq R \leq C(p),$$

where $C(p) = 1 - h_2(p)$ is the BSC capacity, and the functions $E_{\text{sp}}(R, p)$ and $R_{\text{crit}}(p)$ are defined in (6) and (8) below. Moreover, if $p$ is not too small then the formula holds

$$E(R, p) = E_{\text{sp}}(R_{\text{crit}}(p), p) + R_{\text{crit}}(p) - R, \qquad R_1(p) \leq R \leq R_{\text{crit}}(p), \qquad (2)$$

where $R_1(p)$ is defined in (6). Using the method of [9] the formula (2) has been obtained in [6, 7] for $p \geq 0.046$, and then in [8] for $p \geq 0.04468$. A corollary of this paper main result (theorem 1) establishes the formula (2) for $p > p_1 \approx 0.0078176$. Here the value $p_1$ is the minimal possible for validity of (2) (otherwise, $R_1(p) > R_{\text{crit}}(p)$).

For $0 < R < \min\{R_{\text{crit}}(p), R_1(p)\}$ still only some lower and upper bounds for the function $E(R, p)$ are known [1 – 8], and in this paper the most accurate of the upper bounds [8] is improved.

Let us explain what constituted the difficulty in upper bounding the function $E(R, p)$ in the earlier papers [5, 7, 8]. It was shown in [5] that for a $(R, n)$–code there exists $\omega$ such that $B_{\omega n} \gtrsim 2^{\mu(R,\omega)n}$, where the function $\mu(R, \omega) > 0$ is described below, and $\omega$ does not exceed the best upper bound (linear programming) for the minimal code distance $d(\mathcal{C})$ (that result improved the earlier known similar bound in [10]). Note that when testing only two codewords $\boldsymbol{x}_i, \boldsymbol{x}_j$ with large $d(\boldsymbol{x}_i, \boldsymbol{x}_j) = d$ for the decoding error probability $P_{\text{e}}$ we have $P_{\text{e}} \sim (4pq)^{d/2}$. Therefore, if each codeword $\boldsymbol{x}_i$ has approximately $B_{\omega n}$ neighbors on the distance $\omega n$, then it is natural to expect that $P_{\text{e}} \gtrsim (4pq)^{\omega n/2} B_{\omega n}$ for large $n$ (and not very large $\omega$), i.e. some kind of an additive *lower* bound for the probability of the union of events holds.

The first variant of such additive bound was obtained in [5] under rather severe constraints on $R$. Those results of [5] have been strengthened in [6, 7], using the method of [9] (which, in turn, goes to [11 – 13]). There were still a certain constraints on $R, p$. Results of [6, 7] were strengthened in [8] under a milder constraints on $R, p$. It should be noted that in all those papers a various variants of the second order Bonferroni inequalities were used.

The aim of this paper is to prove that additive bound without constraints on $R, p$. For that purpose the earlier method of [8, 9] (see also [11 – 13]) is significantly strengthened (and simplified). Moreover, Bonferroni inequalities are not used. That approach allows to treat in a unified way both a low and a high rate $R$ cases. As an example, in §2 a new derivation of the sphere-packing bound is presented. Proposition 4 plays an important role in that method. It partially answers the natural question: how many Voronoi regions (i.e. how many codewords $\boldsymbol{x}_i$) determine the output $\boldsymbol{y}$ ?) The answer: for the "essential fraction" of all $\boldsymbol{y}$'s the number of such $\boldsymbol{x}_i$ does not exceed $n^2$.

The approach used in the paper is applicable to other channels as well. In particular, for the Gaussian channel it will be done in the next paper.

We describe the main paper results. Introduce the function [14]

$$G(\alpha, \tau) = 2\frac{\alpha(1-\alpha) - \tau(1-\tau)}{1 + 2\sqrt{\tau(1-\tau)}} = \frac{1}{2} - \sqrt{\tau(1-\tau)} - \frac{(1-2\alpha)^2}{2\left[1 + 2\sqrt{\tau(1-\tau)}\right]}. \qquad (3)$$

If $x_{\tau n}^{\alpha n}$ is the minimal root of the Hahn polynomial $Q_{\tau n}^{(n,\alpha n)}(x)$, then $2x_{\tau n}^{\alpha n} = G(\alpha,\tau)n[1 + o(1)]$, $n \to \infty$ [14, formula (B.21)], [15, Corollary 5.22]. It is known that [16, 5, 8]

$$\min_{\substack{0 \leq \tau \leq \alpha \leq 1/2 \\ h_2(\alpha) - h_2(\tau) = 1 - R}} G(\alpha, \tau) = G\left(1/2, h_2^{-1}(R)\right), \tag{4}$$

if and only if

$$R \leq R_0 = h_2(\tau_0) \approx 0.30524, \tag{5}$$

where $\tau_0 \approx 0.054507$ is the unique root of the equation

$$f(\tau) = (1 - 2\tau)\left[1 + \frac{1}{2\sqrt{\tau(1-\tau)}}\right] - \ln\frac{1-\tau}{\tau} = 0.$$

Throughout the paper we use the values and functions

$$\begin{aligned}
\tau_1(p) &= \frac{\left(1 - (4pq)^{1/4}\right)^2}{2(1 + \sqrt{4pq})}, \qquad R_1(p) = h_2(\tau_1(p)), \\
\omega_1(p) &= G(1/2, \tau_1(p)) = \frac{\sqrt{4pq}}{1 + \sqrt{4pq}}, \\
\omega_0(\tau) &= G(1/2, \tau) = \frac{1}{2} - \sqrt{\tau(1-\tau)}, \quad \omega_0(\tau_0) \approx 0.27298, \\
R_{\mathrm{crit}}(p) &= h_2(\tau_{\mathrm{crit}}(p)) = 1 - h_2\left(\frac{\sqrt{p}}{\sqrt{p}+\sqrt{q}}\right).
\end{aligned} \tag{6}$$

We have
$$\begin{aligned}
R_1(p) &\leq R_{\mathrm{crit}}(p), \quad \text{if and only if} \quad p \geq p_1 \approx 0.0078176, \\
\omega_1(p_1) &\approx 0.149762, \quad \tau_1(p_1) \approx 0.1431616, \\
R_{\mathrm{crit}}(p) &\leq R_0, \quad p \geq 0.05014,
\end{aligned} \tag{7}$$

where $p_1$ is the unique root of the equation $R_1(p) = R_{\mathrm{crit}}(p)$.

Remind the sphere-packing upper bound

$$\begin{aligned}
E(R, p) &\leq E_{\mathrm{sp}}(R, p) = D\left(h_2^{-1}(1-R)\|p\right), \quad 0 \leq R \leq C(p), \\
D(x\|y) &= x\log\frac{x}{y} + (1-x)\log\frac{1-x}{1-y}, \\
E_{\mathrm{sp}}(0, p) &= \frac{1}{2}\log\frac{1}{4pq} = 2E(0, p).
\end{aligned} \tag{8}$$

Introduce also the functions

$$W(\omega, \alpha, R, p) = \frac{\omega}{2}\log\frac{1}{4pq} - \mu(R, \alpha, \omega), \tag{9}$$

where $\mu(R, \alpha, \omega)$ is defined in (14), and

$$F_1(R, \alpha, p) = \max_{0 \leq \omega \leq G(\alpha, \tau)} W(\omega, \alpha, R, p), \qquad \tau = h_2^{-1}(h_2(\alpha) - 1 + R) \leq 1/2,$$
$$F(R, p) = \min_{\alpha_0(R) \leq \alpha \leq 1/2} F_1(R, \alpha, p), \qquad \alpha_0(R) = h_2^{-1}(1 - R). \tag{10}$$

The following theorem represents the main paper result. It strengthens a similar Theorem 1 from [8].

T h e o r e m  1. *For any $0 \leq R \leq C(p)$ the upper bound holds*

$$E(R, p) \leq \min\{F(R, p), E_{\text{sp}}(R, p)\}. \tag{11}$$

The best upper bound for $E(R, p)$ is a combination of the inequality (11) with the "straight-line bound" [2]. In particular, using also the random coding lower bound for $E(R, p)$ [1, 3] we get

C o r o l l a r y  1. *If $R_1(p) < R_{\text{crit}}(p)$ or, equivalently, if $p > p_1 \approx 0.0078176$, then*

$$E(R, p) = \begin{cases} 1 - \log_2\left(1 + 2\sqrt{pq}\right) - R, & R_1(p) \leq R \leq R_{\text{crit}}(p), \\ E_{\text{sp}}(R, p), & R_{\text{crit}}(p) \leq R \leq C(p). \end{cases} \tag{12}$$

In the right-hand side of (12) the relation was used (see (2))

$$E_{\text{sp}}(R_{\text{crit}}, p) + R_{\text{crit}} = E(R_1, p) + R_1 = 1 - \log_2\left(1 + 2\sqrt{pq}\right).$$

Therefore for $R \geq \min\{R_{\text{crit}}(p), R_1(p)\}$ (i.e. for $\tau \geq \min\{\tau_{\text{crit}}(p), \tau_1(p)\}$) the function $E(R, p)$ is now known exactly without any additional constraints. Earlier that result was obtained for $p \geq 0.046$ [6, 7], and then for $p \geq 0.04468$ [8]. Corollary 1 establishes that result to the minimal possible $p$.

Note that $\max_p \{R_{\text{crit}}(p) - R_1(p)\} \approx 0.07131$, and it is attained for $p \approx 0.0922$. Moreover, $R_{\text{crit}}(p) \approx 0.20219$, $C(p) \approx 0.5562$. We also have $\lim_{p \to 0} R_{\text{crit}}(p) = \lim_{p \to 0} R_1(p) = 1$.

C o r o l l a r y  2. *For $R \leq \min\{R_{\text{crit}}(p), R_1(p)\}$ optimal $\alpha, \omega$ in (10) are $\alpha = 1/2$, $\omega = G(1/2, \tau) = 1/2\sqrt{\tau(1-\tau)}$, and then*

$$F(R, p) = W(G(1/2, \tau), 1/2, R, p) =$$
$$= \frac{G(1/2, \tau)}{2} \log \frac{1}{4pq} - h_2(\tau) - h_2(G(1/2, \tau)) + 1. \tag{13}$$

The formula (13) follows from (17), (18) and (16).

*Remark* 1. Generally speaking, it is not possible to replace $F(R, p)$ in the right-hand side of (11) on $W(G(1/2, \tau), 1/2, R, p)$ for any $R, p$. In particular, it may be wrong when $p$ is small, and $R$ is large. For example, for $p = 0.001$, $\tau = 0.4$ we have $R \approx 0.971 < C \approx 0.989$, but $W(G(1/2, \tau), 1/2, R, p) \approx -0.0122$.

In §2 the known result on a code spectrum [5] and some analytic properties of the related functions are described. In §3 a new approach, in general, is presented, and, as an example, a new derivation of the sphere-packing bound is presented. In §4 this approach is developed using the code spectrum, and theorem 1 is proved.

## II. Code spectrum and functions analytic properties

To describe the known result on a code spectrum [5] introduce the function

$$\mu(R, \alpha, \omega) = -h_2(\alpha) - 2(1-\omega)\log(1-\omega) - 2\omega \log e +$$
$$+(1-\omega)h_2\left(\frac{\alpha - \omega/2}{1-\omega}\right) - 2\int_0^{\omega/2} \log\left[s(u) + 2u^2 + \sqrt{s^2(u) - 4u^2\tau(1-\tau)}\right] du, \quad (14)$$
$$s(u) = \alpha(1-\alpha) - \tau(1-\tau) - u,$$
$$\tau = h_2^{-1}(h_2(\alpha) - 1 + R) \leq 1/2.$$

The function $\mu(R, \alpha, \omega)$ is a more convenient representation of the same function from [5]. The next result is a variant of [5, theorem 5]. In Appendix its proof is presented since the proof in [5] causes some questions.

T h e o r e m 2. *For any $(R, n)$–code and any $\alpha \in [h_2^{-1}(1-R), 1/2]$ there exists the value $\omega$, $0 \leq \omega \leq G(\alpha, \tau)$ such that $n^{-1} \log B_{\omega n} \geq \mu(R, \alpha, \omega) + o(1)$ as $n \to \infty$, where $\tau = h_2^{-1}(h_2(\alpha) - 1 + R)$, and $G(\alpha, \tau)$ and $\mu(R, \alpha, \omega)$ are defined in (3) and (14).*

The value $\alpha = 1/2$ is the best in Theorem 2 for $0 \leq R \leq R_0$, where $R_0$ is defined in (5) [8, Remark 4]. Since for $\alpha = 1/2$ we have

$$s(u) + 2u^2 + \sqrt{s^2(u) - 4u^2\tau(1-\tau)} = \frac{1}{8}\left[1 - 2\tau + \sqrt{(1-2\tau)^2 - 8u + 16u^2}\right]^2,$$

then for the function $\mu(R, 1/2, \omega)$, $\omega \leq G(1/2, \tau)$, with $\tau = h_2^{-1}(R)$ after integration in (14) we have [8]

$$\mu(R, 1/2, \omega) = -2(1-\omega)\log(1-\omega) - 1 - \left(\frac{3}{2} - 2\tau\right)\log(1-\tau) - \frac{1}{2}\log \tau +$$
$$+(1-2\omega)\log g + (1-2\tau)\log[g+\tau-\omega]] - \frac{1}{2}\log\frac{(1-2\tau)g - \omega}{1 - (1-2\tau)g - \omega}, \quad (15)$$
$$g = g(\tau, \omega) = \frac{1 - 2\tau + \sqrt{(1-2\tau)^2 - 4\omega(1-\omega)}}{2}.$$

In particular, for $\omega = G(1/2, \tau)$ we get from (15)

$$\mu\left(h_2(\tau), 1/2, G(1/2, \tau)\right) = h_2(\tau) + h_2(G(1/2, \tau)) - 1. \quad (16)$$

The next result which is a combination of Propositions 1 and 2 from [8], describes some analytical properties of the functions $W(\omega, \alpha, R, p)$, $F_1(R, \alpha, p)$ and $F(R, p)$ from (9)–(10).

P r o p o s i t i o n 1. 1) *For any $\alpha_0(R) \leq \alpha < 1/2$ and $\omega > 0$ we have*

$$\frac{d\mu(R, \alpha, \omega)}{d\alpha} > 0, \qquad \alpha_0(R) = h_2^{-1}(1-R).$$

2) *For any* $0 \leq \tau \leq \alpha \leq 1/2$ *and* $0 < \omega < G(\alpha, \tau)$ *we have*

$$\mu''_{\omega^2}(R, \alpha, \omega) > 0.$$

*Moreover,* $W'_\omega(\omega, \alpha, R, p)\big|_{\omega=G(\alpha,\tau)} \geq 0$, *and then* $F_1(R, \alpha, p) = W(G(\alpha, \tau), \alpha, R, p)$, *if and only if*

$$G(\alpha, \tau) \geq \frac{2\sqrt{pq}}{1 + 2\sqrt{pq}} = \omega_1(p) = G(1/2, \tau_1(p)).$$

3) *With* $R_0$ *from* (4) *for any* $p$ *we have*

$$F(R, p) = \min_{\alpha_0(R) \leq \alpha \leq 1/2} F_1(R, \alpha, p) = F_1(R, 1/2, p), \qquad 0 \leq R \leq R_0. \qquad (17)$$

In particular, with $\tau_1(p)$ from (6) for any $p$ we have [8, Proposition 3]

$$F_1(R, 1/2, p) = W(G(1/2, \tau), 1/2, R, p), \quad \text{if and only if} \quad \tau \leq \tau_1(p). \qquad (18)$$

### III. New approach and sphere-packing exponent

In the approach used in [8, 9] we evaluated the individual decoding error probabilities $P_e(\boldsymbol{x}_j|\boldsymbol{x}_i)$, $j \neq i$. Instead, it turns out possible (and easier) to evaluate the total decoding error probability $P_e = M^{-1} \sum_i \sum_{j \neq i} P_e(\boldsymbol{x}_j|\boldsymbol{x}_i)$. When doing so the following fact is very important: for an "essential fraction" of all outputs $\boldsymbol{y}$ the number of the closest to $\boldsymbol{y}$ codewords $\boldsymbol{x}_i$ does not exceed $n^2$ (see Lemma 3).

To describe our approach, for an integer $t$ and an output $\boldsymbol{y}$ define the set:

$$\mathbf{X}_t(\boldsymbol{y}) = \{\boldsymbol{x}_i \in \mathcal{C} : d(\boldsymbol{y}, \boldsymbol{x}_i) = t\}, \qquad \boldsymbol{y} \in \mathbf{Y}. \qquad (19)$$

All codewords $\{\boldsymbol{x}_i\}$ are assumed equiprobable. For a chosen decoding method denote $P(e|\boldsymbol{y}, \boldsymbol{x}_i)$ the conditional decoding error probability provided that $\boldsymbol{x}_i$ was transmitted and $\boldsymbol{y}$ was received. Denote $P_e(\boldsymbol{y})$ the probability to get the output $\boldsymbol{y}$ and to make a decoding error. Then we have

$$P_e(\boldsymbol{y}) = \frac{1}{M} \sum_{i=1}^{M} p(\boldsymbol{y}|\boldsymbol{x}_i) P(e|\boldsymbol{y}, \boldsymbol{x}_i) = \frac{1}{M} \sum_{t=0}^{n} \sum_{\boldsymbol{x}_i \in \mathbf{X}_t(\boldsymbol{y})} p(\boldsymbol{y}|\boldsymbol{x}_i) P(e|\boldsymbol{y}, \boldsymbol{x}_i) =$$

$$= \frac{q^n}{M} \sum_{t=0}^{n} \left(\frac{p}{q}\right)^t \sum_{\boldsymbol{x}_i \in \mathbf{X}_t(\boldsymbol{y})} P(e|\boldsymbol{y}, \boldsymbol{x}_i) \geq \frac{q^n}{M} \sum_{t=0}^{n} \left(\frac{p}{q}\right)^t [|\mathbf{X}_t(\boldsymbol{y})| - 1]_+,$$

where $[z]_+ = \max\{0, z\}$. For the decoding error probability $P_e$ we get

$$P_e = \sum_{\boldsymbol{y} \in \mathbf{Y}} P_e(\boldsymbol{y}) \geq \frac{q^n}{M} \sum_{t=0}^{n} \left(\frac{p}{q}\right)^t \sum_{\boldsymbol{y}:|\mathbf{X}_t(\boldsymbol{y})| \geq 2} [|\mathbf{X}_t(\boldsymbol{y})| - 1].$$

Using the simple inequality $(a-1) \geq a/2$, $a \geq 2$, we get

$$P_{\mathrm{e}} \geq \frac{q^n}{2M} \sum_{t=0}^{n} \left(\frac{p}{q}\right)^t \sum_{\boldsymbol{y}:|\mathbf{X}_t(\boldsymbol{y})|\geq 2} |\mathbf{X}_t(\boldsymbol{y})|, \qquad (20)$$

where $\mathbf{X}_t(\boldsymbol{y})$ is defined in (19). To develop the right-hand side of (20) denote

$$\mathbf{Z}_t(i) = \{\boldsymbol{y} : d(\boldsymbol{y}, \boldsymbol{x}_i) = t, \, |\mathbf{X}_t(\boldsymbol{y})| \geq 2\} =$$
$$= \{\boldsymbol{y} : d(\boldsymbol{y}, \boldsymbol{x}_i) = t \text{ and there exists } \boldsymbol{x}_j \neq \boldsymbol{x}_i \text{ with } d(\boldsymbol{x}_j, \boldsymbol{y}) = t\}.$$

Then we can represent the inequality (20) as follows

$$P_{\mathrm{e}} \geq \frac{q^n}{2M} \sum_{t=0}^{n} \left(\frac{p}{q}\right)^t \sum_{i=1}^{M} |\mathbf{Z}_t(i)|. \qquad (21)$$

As a result, we get

P r o p o s i t i o n 2. *For the decoding error probability $P_{\mathrm{e}}$ the lower bounds* (20) *and* (21) *hold.*

*Remark* 2. Although simple, the lower bounds (20) and (21)–(22) give the exact exponent order of the best $P_{\mathrm{e}}$ for any rate $R$. It is so because for the optimal (maximum likelihood) decoding we have equalities (for exponents) in all their derivation steps.

**Example - Sphere-packing upper bound.** We show first how to get the sphere-packing upper bound (8) from (21). Note that

$$\mathbf{Z}_t(i) = \mathbf{Z}_t^{(1)}(i) \setminus \mathbf{Z}_t^{(2)}(i),$$
$$\mathbf{Z}_t^{(1)}(i) = \{\boldsymbol{y} : d(\boldsymbol{y}, \boldsymbol{x}_i) = t\},$$
$$\mathbf{Z}_t^{(2)}(i) = \{\boldsymbol{y} : d(\boldsymbol{y}, \boldsymbol{x}_i) = t, \, |\mathbf{X}_t(\boldsymbol{y})| = 1\}.$$

Therefore we have

$$\sum_{i=1}^{M} |\mathbf{Z}_t(i)| = M \left|\mathbf{Z}_t^{(1)}(1)\right| - |\mathbf{Y}_t| = M\binom{n}{t} - |\mathbf{Y}_t|,$$
$$\mathbf{Y}_t = \{\boldsymbol{y} : |\mathbf{X}_t(\boldsymbol{y})| = 1\} = \qquad (22)$$
$$= \{\boldsymbol{y} : \text{ there exists exactly one } \boldsymbol{x}_i \text{ with } d(\boldsymbol{y}, \boldsymbol{x}_i) = t\}.$$

Since $|\mathbf{Y}_t| \leq 2^n$, we get from (21) and (22)

$$P_{\mathrm{e}} \geq \frac{q^n}{2M} \max_{0 \leq t \leq n} \left\{\left(\frac{p}{q}\right)^t \left[M\binom{n}{t} - 2^n\right]\right\}. \qquad (23)$$

To maximize the right-hand-side of (23) for $M = 2^{Rn}$, $R > 0$, it is sufficient to set $t = x_0 n + \delta$, where

$$R + h_2(x_0) = 1, \qquad \delta = \frac{3\log(n+1)}{\log((1-x_0)/x_0)}.$$

Then for large $n$ we have $M\binom{n}{t} - 2^n \geq 2^n \log(n+1)$, and from (23) we get the sphere-packing upper bound (8).

## IV. Lower bound (21) and code spectrum

To investigate the reliability function $E(R)$ for $R < R_{\text{crit}}$ we use the simplified (but exponentially equivalent) form of the lower bound (21)

$$P_{\text{e}} \geq \frac{q^n}{2M} \max_{0 \leq t \leq 1} \max_\omega \left(\frac{p}{q}\right)^{tn} \sum_{i=1}^{M} |\mathbf{Z}_i(t,\omega)|, \tag{24}$$

where

$$\mathbf{Z}_i(t,\omega) = \{\mathbf{y} : \text{there exists } \mathbf{x}_j \text{ with } d(\mathbf{x}_i, \mathbf{x}_j) = \omega n \text{ and } d(\mathbf{x}_i, \mathbf{y}) = d(\mathbf{x}_j, \mathbf{y}) = tn\}.$$

We develop the lower bound (24), relating it to the code spectrum (1), i.e. to the distribution of the pairwise distances $d(\mathbf{x}_i, \mathbf{x}_j)$. Note that in (24) the sets $\{\mathbf{Z}_i(t,\omega)\}$ can be replaced by any their subsets.

For a codewords $\mathbf{x}_i, \mathbf{x}_j$ with $d(\mathbf{x}_i, \mathbf{x}_j) = d_{ij} = \omega n$ introduce the set

$$\mathbf{Z}_{ij}(t,\omega) = \{\mathbf{y} : d(\mathbf{x}_i, \mathbf{y}) = d(\mathbf{x}_j, \mathbf{y}) = tn\}.$$

Then for any $i$ we have

$$\mathbf{Z}_i(t,\omega) = \bigcup_{j : d_{ij} = \omega n} \mathbf{Z}_{ij}(t,\omega).$$

Denote

$$\begin{aligned}\mathbf{X}_i(\mathbf{y}, t, \omega) &= \{\mathbf{x}_j : d(\mathbf{x}_i, \mathbf{x}_j) = \omega n, \, d(\mathbf{x}_i, \mathbf{y}) = d(\mathbf{x}_j, \mathbf{y}) = tn\}, \\ X_{\max}(t,\omega) &= \max_{i,\mathbf{y}} |\mathbf{X}_i(\mathbf{y}, t, \omega)|.\end{aligned} \tag{25}$$

Note that if $A_1, \ldots, A_N$ are finite sets, and any point $a \in \bigcup_i A_i$ is covered by the sets $\{A_i\}$ not more than $K$ times, then

$$\left|\bigcup_{i=1}^{N} A_i\right| \geq \frac{1}{K} \sum_{i=1}^{N} |A_i|.$$

Therefore we have

$$|\mathbf{Z}_i(t,\omega)| \geq \frac{1}{X_{\max}(t,\omega)} \sum_{j : d_{ij} = \omega n} |\mathbf{Z}_{ij}(t,\omega)|. \tag{26}$$

Since the cardinality $|\mathbf{Z}_{ij}(t,\omega)|$ does not depend on indices $(i,j)$, we denote it simply $Z(t,\omega)$. As a result, we get from (24) and (26)

$$P_{\text{e}} \geq \frac{q^n}{2} \max_{0 \leq t \leq 1} \max_\omega \left\{ \left(\frac{p}{q}\right)^{tn} \frac{B_{\omega n} Z(t,\omega)}{X_{\max}(t,\omega)} \right\}. \tag{27}$$

For the value $Z(t,\omega)$, $\omega/2 \leq t \leq 1/2$ we have $(n \to \infty)$

$$\frac{1}{n}\log_2 Z(t,\omega) = \frac{1}{n}\log_2\left[\binom{(1-\omega)n}{(t-\omega/2)n}\binom{\omega n}{\omega n/2}\right] = u(t,\omega) + o(1),$$

$$u(t,\omega) = \omega + (1-\omega)h_2\left(\frac{1}{2} - \frac{1-2t}{2(1-\omega)}\right), \quad (28)$$

$$u'_\omega(t,\omega) = \frac{1}{2}\log_2\left[1 - \frac{(1-2t)^2}{(1-\omega)^2}\right] \leq 0.$$

We formulate the result obtained as follows.

P r o p o s i t i o n 3. *For the decoding error probability $P_e$ the lower bound (27) holds, where $X_{\max}(t,\omega)$ and $Z(t,\omega)$ are defined in (25) and (28), respectively.*

We set below

$$t(\omega) = \min\left\{\frac{\omega}{2} + (1-\omega)p, \frac{1-\sqrt{1-2\omega}}{2}\right\} =$$

$$= \begin{cases} \omega/2 + (1-\omega)p, & \omega \geq \omega_1(p), \\ \left(1 - \sqrt{1-2\omega}\right)/2, & \omega \leq \omega_1(p), \end{cases} \quad (29)$$

where $\omega_1(p)$ is defined in (6). Such choice of the function $t(\omega)$ has a certain optimal properties (see Remark 4 further). In particular, after a "cleaning" procedure [8] the value $X_{\max}(t(\omega),\omega)$ does not exceed $n^2$, and then it will not influence the exponent of $P_e$.

To upper bound the value $X_{\max}(t(\omega),\omega)$ we use the following known result. Let $A(n,d,w)$ be the maximal cardinality of a constant weight $w$ code $\{\boldsymbol{x}_i\}$ with $d(\boldsymbol{x}_i,\boldsymbol{x}_j) \geq d$, $i \neq j$. Due to Johnson's bound [17, Theorem 17.2.2] and [17, Theorem 17.2.4] we have

$$A(n,d,w) \leq \frac{dn}{(2w^2 - 2wn + dn)_+}, \quad (30)$$

$$A(n,d,w) \leq \frac{n}{w}A(n-1,d,w-1). \quad (31)$$

Consider first

**The model case**. It exhibits an important property of the function $t(\omega)$ from (29).

P r o p o s i t i o n 4. *Let $\{\boldsymbol{x}_1,\ldots,\boldsymbol{x}_A\}$ be a code of length $n$ with $w(\boldsymbol{x}_i) = t(\omega)n$ and $d(\boldsymbol{x}_i,\boldsymbol{x}_j) \geq \omega n$, $i \neq j$, and the function $t(\omega)$ be defined in (29). Then for any $\omega \leq 1/2$ and $p$ the inequality holds*

$$A < n^2. \quad (32)$$

P r o o f. If $\omega \neq \omega_1(p)$, then with $t(\omega)$ from (29) we have $2w^2 - 2wn + dn > 0$, and hence $2w^2 - 2wn + dn \geq 1$ (since the last value is an integer). Therefore from (30) we get (32). If $\omega = \omega_1(p)$, then $2w^2 - 2wn + dn = 0$. Now we use first the inequality (31), and then to the value $A(n-1,d,w-1)$ apply the upper bound (30). Then with $w = t(\omega_1)n$ and $d = \omega_1 n$ we have

$$2(w-1)^2 - 2(w-1)(n-1) + d(n-1) = 2n - 2w - d =$$
$$= n[2 - 2t(\omega_1) - \omega_1] = 2n(1-\omega_1)q.$$

Therefore from (30) and (31) we get

$$A(n,d,w) \leq \frac{n}{w}A(n-1,d,w-1) \leq \frac{d(n-1)}{2(1-\omega_1)qw} =$$
$$= \frac{\omega_1(n-1)}{2q(1-\omega_1)t(\omega_1)} = \frac{(n-1)}{\sqrt{q}} \leq n\sqrt{2},$$

from which we again get (32). ▲

For a code $\{\boldsymbol{x}_1, \ldots, \boldsymbol{x}_M\}$ consider, say, the codeword $\boldsymbol{x}_1$ and its neighbors $\{\boldsymbol{x}_i\}$ with $d(\boldsymbol{x}_1, \boldsymbol{x}_i) = \omega n$. Suppose that all those neighbors $\{\boldsymbol{x}_i\}$ satisfy also the condition $d(\boldsymbol{x}_i, \boldsymbol{x}_j) \geq \omega n$, $i \neq j$. Due to (32) for any $\boldsymbol{y} \in \bigcup_i \mathbf{Z}_{1i}(t(\omega), \omega)$ there exist less than $n^2$ neighbors $\{\boldsymbol{x}_i\}$, satisfying the inequality

$$d(\boldsymbol{x}_1, \boldsymbol{y}) = d(\boldsymbol{x}_i, \boldsymbol{y}) = t(\omega)n,$$

i.e. in that model case we have $|\mathbf{X}_1(\boldsymbol{y}, t(\omega), \omega)| < n^2$ for any $\boldsymbol{y}$.

Assume now that for the whole code $\{\boldsymbol{x}_1, \ldots, \boldsymbol{x}_M\}$ we have $d(\boldsymbol{x}_i, \boldsymbol{x}_j) \geq \omega n$, $i \neq j$. Then $|\mathbf{X}_i(\boldsymbol{y}, t(\omega), \omega)| < n^2$ for any $i$ and $\boldsymbol{y}$, i.e. $X_{\max}(t(\omega), \omega) < n^2$. Therefore in such a model case $X_{\max}(t(\omega), \omega)$ does not influence the exponent of the value $P_{\mathrm{e}}$.

*Remark* 3. In essence, proposition 4 asserts that for an "essential fraction" of all $\boldsymbol{y}$ the number of Voronoi regions determining $\boldsymbol{y}$, does not exceed $n^2$.

**General case**. In the general case for some $\omega$ we are interested in a pairs $(\boldsymbol{x}_i, \boldsymbol{x}_j)$ with $d_{ij} = \omega n$. But there may exist a pairs $(\boldsymbol{x}_k, \boldsymbol{x}_l)$ with $d_{kl} < \omega n$. Using the "cleaning" procedure [8] we show that the influence of such pairs $(\boldsymbol{x}_k, \boldsymbol{x}_l)$ on the value $P_{\mathrm{e}}$ is not large. It will allow us to reduce the general case to the model case,

Remind that $MB_{vn}$ is the total number of pairs $(\boldsymbol{x}_i, \boldsymbol{x}_j)$ with $d_{ij} = vn$, and $Z(t(v), v)$ for such pair $(\boldsymbol{x}_i, \boldsymbol{x}_j)$ is the number of points $\boldsymbol{y}$ such that

$$d(\boldsymbol{x}_i, \boldsymbol{y}) = d(\boldsymbol{x}_j, \boldsymbol{y}) = t(v)n.$$

Consider the function

$$\left(\frac{p}{q}\right)^{t(v)n} B_{vn} Z(t(v), v), \qquad v \leq G(\alpha, \tau), \tag{33}$$

and assume that it attains its maximum for some for $v = s \leq G(\alpha, \tau)$ (if there are several such $s$, we choose the smallest one). We show that for large $n$ the inequality holds

$$\frac{1}{n}\log\frac{1}{P_{\mathrm{e}}} \leq -\frac{1}{n}\log\left[q^n \left(\frac{p}{q}\right)^{t(s)n} B_{sn} Z(t(s), s)\right] \leq$$
$$\leq \max_{s \leq G(\alpha,\tau)} \left\{\frac{s}{2}\log\frac{1}{4pq} - \mu(R, \alpha, s)\right\} = F_1(R, \alpha, p). \tag{34}$$

Then from (34) and (10) the upper bound (11) and theorem 1 will follow.

To make the cleaning procedure we estimate the value $B_{\lambda n} Z(t(s), \lambda)$, $\lambda < s$. By definition of the value $s$ for any $\lambda \neq s$ we have

$$B_{\lambda n} < \left(\frac{p}{q}\right)^{t(s)n - t(\lambda)n} B_{sn} \frac{Z(t(s), s)}{Z(t(\lambda), \lambda)}.$$

Multiplying both sides of that formula by $Z(t(s), \lambda)$, due to (28) we get that it is equivalent to the inequality

$$\frac{1}{n} \log \frac{B_{\lambda n} Z(t(s), \lambda)}{B_{sn} Z(t(s), s)} < g(\lambda, s),$$

$$g(\lambda, s) = u(t(s), \lambda) - u(t(\lambda), \lambda) - [t(s) - t(\lambda)] \log \frac{q}{p} = \qquad (35)$$

$$= (1 - \lambda) \left[ h_2 \left( \frac{1}{2} - \frac{1 - 2t(s)}{2(1 - \lambda)} \right) - h_2 \left( \frac{1}{2} - \frac{1 - 2t(\lambda)}{2(1 - \lambda)} \right) \right] - [t(s) - t(\lambda)] \log \frac{q}{p}.$$

Depending on $\omega$ the function $t(\omega)$ has one of two forms (see the formula (29)). Therefore there are possible three location cases of the values $\lambda < s$ and $\omega_1(p)$: 1) $\lambda < s \leq \omega_1(p)$, 2) $\omega_1(p) \leq \lambda < s$, and 3) $\lambda < \omega_1(p) < s$. We consider them separately.

1) *Case $\lambda < s \leq \omega_1(p)$.* Essentially, it corresponds to the case $R \geq R_1(p)$. Since $R_1(p) < R_{\text{crit}}(p)$, if $p > p_1 \approx 0.0078176$, we may assume that $p > p_1$. It is sufficient to consider the case $R = R_1(p)$, and after that for $R_1(p) < R < R_{\text{crit}}(p)$ to use the "straight-line bound". For that purpose we prove the following result.

P r o p o s i t i o n 5. *If $p > p_1 \approx 0.0078176$ (see (7)), then*

$$E(R_1(p), p) \leq \min_{\alpha} \max_{\omega \leq G(\alpha, \tau_1)} \left\{ \frac{\omega}{2} \log \frac{1}{4pq} - \mu(R_1, \alpha, \omega) \right\} = F(R_1(p), p) = \qquad (36)$$

$$= \frac{\omega_1}{2} \log \frac{1}{4pq} - \mu(R_1, 1/2, \omega_1) = 1 - \log_2 (1 + 2\sqrt{pq}) - R_1(p),$$

where $\omega_1 = G(1/2, \tau_1(p))$, $R_1 = h_2(\tau_1(p))$.

P r o o f. Note that for $p \geq 0.04468$ validity of the relation (36) was proved in [8, corollary 2]. Let $d_{\min} = \omega_{\min} n$ be the minimal code distance. Then we should have

$$\frac{\omega_{\min}}{2} \log \frac{1}{4pq} \geq \frac{\omega_1}{2} \log \frac{1}{4pq} - \mu(R_1, 1/2, \omega_1),$$

(otherwise the desired bound (36) is valid), or, equivalently,

$$\omega_{\min} \geq \omega_1 - \frac{2\mu(R_1, 1/2, \omega_1)}{\log(1/(4pq))} =$$

$$= \frac{2}{\log(1/(4pq))} [1 - \log_2 (1 + 2\sqrt{pq}) - h_2(\tau_1(p))] = \omega_m, \qquad (37)$$

where the formula (16) was used.

For the function $g(\lambda, s)$ from (35) we have

$$
\begin{aligned}
g(\lambda, s) &= (1-\lambda)\left[h_2\left(\frac{1}{2} - \frac{\sqrt{1-2s}}{2(1-s)}\right) - h_2\left(\frac{1}{2} - \frac{\sqrt{1-2\lambda}}{2(1-\lambda)}\right)\right] + \\
&\quad + \frac{1}{2}\left[\sqrt{1-2s} - \sqrt{1-2\lambda}\right]\log\frac{q}{p}, \\
g'_\lambda &= h_2\left(\frac{1}{2} - \frac{\sqrt{1-2\lambda}}{2(1-\lambda)}\right) - h_2\left(\frac{1}{2} - \frac{\sqrt{1-2s}}{2(1-s)}\right) + \\
&\quad + \frac{1}{2\sqrt{1-2\lambda}}\left[\log\frac{q}{p} - \frac{\lambda}{(1-\lambda)}\log\frac{1-\lambda+\sqrt{1-2\lambda}}{1-\lambda-\sqrt{1-2\lambda}}\right] \geq \\
&\geq g'_\lambda\Big|_{s=\omega_1} = h_2\left(\frac{1}{2} - \frac{\sqrt{1-2\lambda}}{2(1-\lambda)}\right) - h_2(p) + \\
&\quad + \frac{1}{2\sqrt{1-2\lambda}}\left[\log\frac{q}{p} - \frac{\lambda}{(1-\lambda)}\log\frac{1-\lambda+\sqrt{1-2\lambda}}{1-\lambda-\sqrt{1-2\lambda}}\right].
\end{aligned}
\tag{38}
$$

Due to (37) we may assume that

$$\omega_m \leq \lambda < s \leq \omega_1.$$

Combining analytical and numerical methods we can check that

$$
\begin{aligned}
\min_{\omega_m \leq \lambda \leq s \leq \omega_1} g''_{\lambda^2}(\lambda, s) &> B(p) > 0, \quad p \geq 0.003, \\
g(\omega_m, \omega_1) &< -0.0008, \quad p \leq 0.22, \\
\max_p \{\omega_1(p) - \omega_m(p)\} &\approx 0.1076,
\end{aligned}
\tag{39}
$$

where the function $B(p)$ monotone decreases with $p$, and, for example, $B(0.003) \approx 0.009$. Since $g(s, s) = 0$, we get from (35) and (38)–(39) for $0.003 \leq p \leq 0.22$ and $\omega_m \leq \lambda < s \leq \omega_1$

$$\frac{1}{n}\log\frac{B_{\lambda n}Z(t(s), \lambda)}{B_{sn}Z(t(s), s)} < g(\lambda, s) < -(s-\lambda)D, \quad D \approx 0.013 \tag{40}$$

Using (40) we get for any $\delta > 0$ and $\omega_m < s \leq \omega_1$

$$\sum_{\omega_m \leq \lambda \leq s-\delta} \frac{B_{\lambda n}Z(t(s), \lambda)}{B_{sn}Z(t(s), s)} < \sum_{\omega_m \leq \lambda \leq s-\delta} 2^{-n(s-\lambda)D} < n2^{-\delta n D}.$$

We set

$$\delta = \delta_1 = \frac{2\log n}{Dn},$$

and then get

$$\sum_{\omega_m \leq \lambda \leq s-\delta_1} B_{\lambda n}Z(t(s), \lambda) < n^{-1}B_{sn}Z(t(s), s), \quad \omega_m \leq s \leq \omega_1(p). \tag{41}$$

Now we perform a cleaning procedure. Consider a pair $(\boldsymbol{x}_i, \boldsymbol{x}_j)$ with $d_{ij} = sn$. We say that such pair $(t,s)$–covers a point $\boldsymbol{y}$ if $d(\boldsymbol{x}_i, \boldsymbol{y}) = d(\boldsymbol{x}_j, \boldsymbol{y}) = tn$. Then $MB_{un}$ pairs $(\boldsymbol{x}_i, \boldsymbol{x}_j)$ with $d_{ij} = sn$ $(t(s), s)$–cover $MB_{sn}Z(t(s), s)$ points $\boldsymbol{y}$ (counting also the covering multiplicities). We disregard all points $\boldsymbol{y}$ that are $(t(s), \lambda)$–covered for $\lambda \leq s - \delta_1$. Then we remain with points $\boldsymbol{y}$ that can be only $(t(s), v)$–covered with $v > s - \delta_1$. Introduce the set

$$\mathbf{Z}'(s, \delta) = \left\{ \boldsymbol{y} : \begin{array}{c} \boldsymbol{y} \text{ is } (t(s), s)\text{-covered and is not} \\ (t(s), \lambda)\text{–covered for any } \lambda \leq s - \delta \end{array} \right\}. \tag{42}$$

Then we have

$$\mathbf{Z}'(s, \delta_1) \subseteq \bigcup_{i=1}^{M} \mathbf{Z}_i(t(s), s). \tag{43}$$

Due to the relation (41) all points $\boldsymbol{y}$ of the set $\mathbf{Z}'(s, \delta_1)$ in total are $(t(s), s)$-covered, at least, $M(1 - 1/n)B_{sn}Z(t(s), s)$ times. For any $\boldsymbol{y} \in \mathbf{Z}'(s, \delta_1)$ and any covering pair $(\boldsymbol{x}_i, \boldsymbol{x}_j)$ we have $d_{ij} \geq (s - \delta_1)n$. Due to the formula (32) the covering multiplicity of any point $\boldsymbol{y} \in \mathbf{Z}'(s, \delta_1)$ does not exceed $n^2$. Therefore for $s \leq \omega_1(p)$ we get from (43)

$$|\mathbf{Z}'(s, \delta_1)| \geq n^{-2} M(1 - 1/n) B_{sn} Z(t(s), s). \tag{44}$$

Now from (44) and (24) we get for $s \leq \omega_1(p)$

$$P_{\text{e}} \geq \frac{q^n (1 - 1/n)}{2n^2} \left( \frac{p}{q} \right)^{t(s)n} B_{sn} Z(t(s), s). \tag{45}$$

Due to the Theorem 2 there exists $\omega \leq G(\alpha, \tau)$ such that $B_{\omega n} \geq 2^{\mu(R, \alpha, \omega)n + o(n)}$. Then by definition of the value $s$ from (33) and (45) we get

$$P_{\text{e}} \geq \frac{q^n (1 - 1/n)}{2n^2} \left( \frac{p}{q} \right)^{t(\omega)n} B_{\omega n} Z(t(\omega), \omega), \qquad 0.003 \leq p \leq 0.22.$$

From the last relation the bound (36) for $0.003 \leq p \leq 0.22$ follows. But that bound is interesting only for $p > p_1 \approx 0.0078176$ (for $p < p_1$ the sphere-packing bound is more accurate).

If $p > 0.22$ then we should perform a cleaning procedure more accurately. We disregard the point $\boldsymbol{y}$ if it is $(t(s), \lambda)$–covered by some pair $(\boldsymbol{x}_i, \boldsymbol{x}_j)$ with $d_{ij} = \lambda n$, where $\lambda \leq s - \delta_1$. Moreover, we shall demand that there exists $\boldsymbol{x}_k$ with $d_{ik} = d_{jk} = sn$ and the point $\boldsymbol{y}$ is also $(t(s), s)$–covered by both pairs $(\boldsymbol{x}_i, \boldsymbol{x}_k)$ and $(\boldsymbol{x}_j, \boldsymbol{x}_k)$. Then we shall essentially follow the computations of [8], introducing the function $m(\omega, \lambda)$, etc. We omit such repeating of the part of the paper [8], moreover, for such $p$ such result was proved in [8]. ▲

2) *Case* $\omega_1(p) \leq \lambda < s$. Taking into account a different form of the function $t(s)$ from

(29), we have

$$g(\lambda, s) = (1 - \lambda) \left[ h_2 \left( \frac{1}{2} - \frac{(1 - 2p)(1 - s)}{2(1 - \lambda)} \right) - h_2(p) \right] - \left( \frac{1}{2} - p \right)(s - \lambda) \log \frac{q}{p},$$

$$g'_\lambda = \frac{1}{2} \log \frac{1}{4pq} + \frac{1}{2} \log \left[ 1 - \frac{(1 - 2p)^2(1 - s)^2}{(1 - \lambda)^2} \right] > 0, \qquad \lambda < s,$$

$$g''_{\lambda\lambda} = -\frac{(1 - 2p)^2(1 - s)^2 \log_2 e}{(1 - \lambda)[(1 - \lambda)^2 - (1 - 2p)^2(1 - s)^2]} < -\frac{(1 - 2p)^2}{3}.$$

Since $g(s, s) = g'_\lambda(s, s) = 0$, we have

$$g(\lambda, s) < -\frac{(1 - 2p)^2}{6}(s - \lambda)^2, \qquad \omega_1(p) \leq \lambda < s,$$

and then for $\omega_1(p) \leq \lambda < s$ we get

$$\frac{1}{n} \log \frac{B_{\lambda n} Z(t(s), \lambda)}{B_{sn} Z(t(s), s)} < -\frac{(1 - 2p)^2}{6}(s - \lambda)^2, \qquad \lambda < s.$$

Now for $\delta > 0$ we have

$$\sum_{\omega_1(p) \leq \lambda \leq s - \delta} \frac{B_{\lambda n} Z(t(s), \lambda)}{B_{sn} Z(t(s), s)} < \sum_{\omega_1(p) \leq \lambda \leq s - \delta} 2^{-(1 - 2p)^2 (s - \lambda)^2 n/6} < n 2^{-(1 - 2p)^2 \delta^2 n/6}.$$

We set

$$\delta = \delta_2 = \frac{\sqrt{12 \log n}}{(1 - 2p)\sqrt{n}},$$

and then get

$$\sum_{\omega_1(p) \leq \lambda \leq s - \delta_2} B_{\lambda n} Z(t(s), \lambda) < n^{-1} B_{sn} Z(t(s), s).$$

Similar to the previous case, we disregard all points $\boldsymbol{y}$ which are $(t(s), \lambda)$–covered for $\lambda \leq s - \delta_2$. Then we get the set $\mathbf{Z}'(s, \delta_2)$ from (42), for which the relation (43) holds. All points $\boldsymbol{y}$ of the set $\mathbf{Z}'(s, \delta_2)$ in total are $(t(s), s)$-covered, at least, $M(1 - 1/n) B_{sn} Z(t(s), s)$ times. For any such covering pair $(\boldsymbol{x}_i, \boldsymbol{x}_j)$ we have $d_{ij} \geq (s - \delta_2)n$. Due to Johnson's bound (30) the covering multiplicity of any point $\boldsymbol{y}$ does not exceed $n^2$, if

$$s > \omega_1(p) + \frac{\delta_2}{\sqrt{4pq + \delta_2(1 - 4pq)} + \sqrt{4pq}}. \tag{46}$$

Therefore for $s$, satisfying the condition (46), we get from (43)

$$|\mathbf{Z}'(s, \delta_2)| \geq n^{-2} M(1 - 1/n) B_{sn} Z(t(s), s). \tag{47}$$

Now from (44) and (24) we again get (45), from which the upper bound (34) follows.

3) *Case* $\lambda < \omega_1(p) < s$. For the function $g(\lambda, s)$ from (35) we use the representation

$$g(\lambda, s) = f(\lambda, s) + g(\lambda, \omega_1(p)),$$
$$f(\lambda, s) = u(t(s), \lambda) - u(t(\omega_1), \lambda) - [t(s) - t(\omega_1)] \log \frac{q}{p},$$

and get

$$f(\lambda, s) = (1 - \lambda) \left[ h_2 \left( \frac{1}{2} - \frac{(1-s)(1-2p)}{2(1-\lambda)} \right) - h_2 \left( \frac{1}{2} - \frac{(1-\omega_1)(1-2p)}{2(1-\lambda)} \right) \right] -$$
$$- \frac{(s - \omega_1)(1 - 2p)}{2} \log \frac{q}{p},$$
$$f'_s(\lambda, s) = \frac{(1 - 2p)}{2} \left[ \log \frac{1 - \lambda + (1-s)(1-2p)}{1 - \lambda - (1-s)(1-2p)} - \log \frac{q}{p} \right] < 0, \qquad \lambda < u,$$
$$f''_{ss}(\lambda, s) = -\frac{(1 - 2p)^2 (1 - \lambda)}{(1 - \lambda)^2 - (1-s)^2 (1-2p)^2} < -(1 - 2p)^2.$$

Therefore from Taylor's formula we have

$$f(\lambda, s) \leq -(1 - 2p)^2 (s - \omega_1)^2,$$

and then, using also (40), we get for $0.003 \leq p \leq 0.22$

$$g(\lambda, s) \leq -(1 - 2p)^2 (s - \omega_1)^2 - (\omega_1 - \lambda) D, \qquad \omega_m \leq \lambda < \omega_1 < s. \tag{48}$$

Now we repeat the same arguments as in the case $\omega_m \leq \lambda < s \leq \omega_1$, but using the inequality (48) instead of (40). As a result, we again get the upper bound (34). It completes the proof of Theorem 1 for $0.003 \leq p \leq 0.22$. For $p > 0.22$ that result was proved in [8]. ▲

*Remark* 4. We explain the choice of the function $t(\omega)$ from (29). Note that when testing only two codewords $(\boldsymbol{x}_i, \boldsymbol{x}_j)$ with large $d_{ij} = \omega n$ for the decoding error probability $P_e$ we have $P_e \sim (4pq)^{\omega n/2}$. Minimal ambiguity set $Y_{ij}$ (i.e. all outputs $\boldsymbol{y}$ for which $\boldsymbol{x}_i, \boldsymbol{x}_j$ are approximately equal) of the same probability has the cardinality of the order $2^{\omega n/2 + (1-\omega) h(p) n}$. If $\omega \geq \omega_1(p)$ (i.e. $R \leq R_1(p)$), then the function $t(\omega)$ from (29) gives exactly such order. In the case $\omega < \omega_1(p)$ (which corresponds to $R > \min\{R_1(p), R_{\text{crit}}(p)\}$) the function $t(\omega)$ in (29) is chosen such that in the right-hand side of (30) to have $2w^2 - 2wn + dn > 0$.

The author thanks L.A.Bassalygo for useful discussions and constructive critical remarks.

# APPENDIX

P r o o f  o f  T h e o r e m  2. For a set $A \subseteq F^n$ and an integer $w$ denote

$$A^{(w)} = \{\boldsymbol{x} \in A : \|\boldsymbol{x}\| = w\}.$$

For some $w \leq n/2$ consider a code $\mathcal{C}^{(w)} = \mathcal{C}(n,w)$ of length $n$ and constant weight $w$. Denote by $B_i^{(w)}$ its spectrum values defined in (1). Using *Hahn polynomials* $Q_j(i) = Q_j^{(n,w)}(i)$ define the polynomials

$$f(x) = \sum_{j=0}^{w} f_j Q_j(x), \tag{A.1}$$

where $f_0 > 0$ and $f_i \geq 0$, $i = 1, \ldots, w$. Then for the code $\mathcal{C}^{(w)}$ the inequality holds

$$\left|\mathcal{C}^{(w)}\right| f(0) + \sum_{\substack{\boldsymbol{x},\boldsymbol{y} \in \mathcal{C}^{(w)} \\ \boldsymbol{x} \neq \boldsymbol{y}}} f(d(\boldsymbol{x},\boldsymbol{y})) \geq \left|\mathcal{C}^{(w)}\right|^2 f_0, \tag{A.2}$$

which follows from [18, formula (1.7) and §6.4]. It follows from (A.2) that

**L e m m a 1.** *For $f(x)$ from (A.1) introduce the set*

$$I = I(f_0, f_1, \ldots, f_w) = \{i \in \{1, \ldots, w\} : f(i) > 0\}. \tag{A.3}$$

*If $d\left(\mathcal{C}^{(w)}\right) > 0$ and $|I| > 0$ then there exists $i \in I$ such that*

$$B_{2i}^{(w)} \geq \frac{f_0 |\mathcal{C}^{(w)}| - f(0)}{|I| f(i)}. \tag{A.4}$$

*Remark* 5. In [5, Lemma 3] the lower bound (A.4) is derived from Delsarte [19] inequalities

$$\sum_{i=0}^{w} B_{2i}^{(w)} Q_j(i) \geq 0, \qquad 0 \leq j \leq w.$$

Next result is a part of derivation of Bassalygo–Elias Lemma (cf. [5, Lemma 2]).

**L e m m a 2.** *For any $w \leq n$ and any $2i \leq \min\{n, 2(n-w)\}$ the relation holds*

$$B_{2i}(\mathcal{C})|\mathcal{C}|\binom{2i}{i}\binom{n-2i}{w-i} = \sum_{\boldsymbol{x} \in F^n} B_{2i}\left((\mathcal{C}+\boldsymbol{x})^{(w)}\right)\left|(\mathcal{C}+\boldsymbol{x})^{(w)}\right|. \tag{A.5}$$

Next result follows from (A.5) and (A.4), and in such form appeared first in [5, Lemma 4].

**L e m m a 3.** *If for $f(x)$ from (A.1) and $I(f)$ from (A.3) we have $|I(f)| > 0$, then for any $w \leq n$ there exists $i \in I(f)$ such that*

$$B_{2i}(\mathcal{C})\binom{2i}{i}\binom{n-2i}{w-i}f(i)|I| \geq \binom{n}{w}\left[f_0 2^{-n}\binom{n}{w}|\mathcal{C}| - f(0)\right]. \tag{A.6}$$

*Proof.* From the formula (A.5) and using (A.4) for some $i \in I$ we have

$$B_{2i}(\mathcal{C})\binom{2i}{i}\binom{n-2i}{w-i}f(i)|I| = \frac{f(i)|I|}{|\mathcal{C}|}\sum_{\boldsymbol{x}\in F^n} B_{2i}\left((\mathcal{C}+\boldsymbol{x})^{(w)}\right)\left|(\mathcal{C}+\boldsymbol{x})^{(w)}\right| \geq$$

$$\geq \frac{1}{|\mathcal{C}|}\sum_{\boldsymbol{x}\in F^n}\left[f_0\left|(\mathcal{C}+\boldsymbol{x})^{(w)}\right|^2 - f(0)\left|(\mathcal{C}+\boldsymbol{x})^{(w)}\right|\right] \geq$$

$$\geq \frac{f_0}{|F^n||\mathcal{C}|}\left[\sum_{\boldsymbol{x}\in F^n}\left|(\mathcal{C}+\boldsymbol{x})^{(w)}\right|\right]^2 - \frac{f(0)}{|\mathcal{C}|}\sum_{\boldsymbol{x}\in F^n}\left|(\mathcal{C}+\boldsymbol{x})^{(w)}\right| =$$

$$= \binom{n}{w}\left[f_0 2^{-n}\binom{n}{w}|\mathcal{C}| - f(0)\right],$$

where at the end we used the Cauchy–Bounjakowsky–Schwarz inequality, and the formula $\sum_{\boldsymbol{x}\in F^n}\left|(\mathcal{C}+\boldsymbol{x})^{(w)}\right| = \binom{n}{w}|\mathcal{C}|$. ▲

Denote $x_t^w$ the minimal root of the polynomial $Q_t^{(n,w)}(x)$. Then $x_{t+1}^w < x_t^w$. We use in (A.6) the same polynomial $f(x)$ as in [14, formulas (4.4) and (4.6)]:

$$f(x) = \frac{1}{(a-x)}Q_t^2(a)\left[Q_t(x) + Q_{t+1}(x)\right]^2, \qquad (A.7)$$

where the value $t \leq w$ will be chosen later, and the parameter $a \in (x_{t+1}^w, x_t^w)$ is such that $Q_t(a) = -Q_{t+1}(a)$. Then $f(a) = 0$ and $f(x) \leq 0$, $a < x \leq w$. Expansion coefficients of $f(x)$ in the polynomials $\{Q_j\}$ basis are nonnegative (see [14]). We also have (see [14])

$$\begin{aligned}f_0 &= \left(\binom{n}{t} - \binom{n}{t-1}\right)\frac{(n-2t)(n-2t-1)}{(t+1)(w-t)(n-w-t)}Q_t^2(a),\\ f(0) &= \frac{1}{a}\left(\binom{n}{t+1} - \binom{n}{t-1}\right)^2 Q_t^2(a).\end{aligned} \qquad (A.8)$$

We choose $t$ such that

$$\frac{f(0)}{f_0} \leq 2^{-(n+1)}\binom{n}{w}|\mathcal{C}|.$$

Then

$$f_0 2^{-n}\binom{n}{w}|\mathcal{C}| - f(0) \geq f_0 2^{-(n+1)}\binom{n}{w}|\mathcal{C}|,$$

and from (A.6) we get

$$B_{2i}(\mathcal{C})\binom{2i}{i}\binom{n-2i}{w-i}f(i)|I(f)| \geq f_0 2^{-(n+1)}|\mathcal{C}|\binom{n}{w}^2.$$

The set $I(f)$ from (A.3) has the form

$$I = I(f) = \{i \in \{0, 1, \ldots, w\} : f(i) > 0\} = \{0, 1, \ldots, a-1\}.$$

Then the bound (A.6) takes the form

$$B_{2i}(\mathcal{C}) \geq \frac{f_0 2^{-(n+1)} |\mathcal{C}| \binom{n}{w}^2}{x_t^w \binom{2i}{i} \binom{n-2i}{w-i} f(i)}, \quad (A.9)$$

and it remains us to evaluate the value $f(i)$. From (A.7) and (A.8) we have

$$\frac{f_0}{f(i)} = \binom{n}{t} \frac{(n-2t+1)(n-2t)(n-2t-1)(a-i)}{(n-t+1)(t+1)(w-t)(n-w-t)\left[Q_t(i) + Q_{t+1}(i)\right]^2}$$

We consider the asymptotic case when $w = \alpha n$, $i = \xi n$, $t = \tau n$, where $\tau \leq \alpha \leq 1/2$ $n \to \infty$. It is known [18] that

$$\frac{x_{\tau n}^{\alpha n}}{n} = \frac{G(\alpha, \tau)}{2} + o(1) = \frac{\alpha(1-\alpha) - \tau(1-\tau)}{1 + 2\sqrt{\tau(1-\tau)}} + o(1).$$

Denote

$$q(\alpha, \tau, \xi) = \lim_{n \to \infty} \frac{1}{n} \log Q_{\tau n}^{(n, \alpha n)}(\xi n). \quad (A.10)$$

Then

$$\lim_{n \to \infty} \frac{1}{n} \log \frac{f_0}{f(i)} = h_2(\tau) - 2q(\alpha, \tau, \xi),$$

and from (A.9) we get (since $h_2(\alpha) - h_2(\tau) = 1 - R$)

$$\lim_{n \to \infty} \frac{1}{n} \log B_{2\xi n}(\mathcal{C}) \geq$$

$$\geq R + 2h_2(\alpha) + h_2(\tau) - 1 - 2\xi - (1-2\xi)h_2\left(\frac{\alpha - \xi}{1 - 2\xi}\right) - 2q(\alpha, \tau, \xi) = \quad (A.11)$$

$$= h_2(\alpha) + 2h_2(\tau) - 2\xi - (1-2\xi)h_2\left(\frac{\alpha - \xi}{1 - 2\xi}\right) - 2q(\alpha, \tau, \xi).$$

Therefore it remains us to evaluate the function $q(\alpha, \tau, \xi)$, moreover it is sufficient to upper bound it (as accurate as possible). How to do that, essentially, is described in [14]. Denote

$$q_0(\alpha, \tau, \xi) = h_2(\tau) - \alpha h_2\left(\frac{\xi}{\alpha}\right) - (1-\alpha)h_2\left(\frac{\xi}{1-\alpha}\right) - 2\xi \log \frac{\xi}{e} - \xi +$$

$$+ \int_0^\xi \log\left[s(u) + 2u^2 + \sqrt{s^2(u) - 4\tau(1-\tau)u^2}\right] du, \quad (A.12)$$

$$s(u) = \alpha(1-\alpha) - \tau(1-\tau) - u.$$

L e m m a 4. *The estimate holds*

$$q(\alpha, \tau, \xi) \leq q_0(\alpha, \tau, \xi), \quad \xi < \frac{\alpha(1-\alpha) - \tau(1-\tau)}{1 + 2\sqrt{\tau(1-\tau)}} = \frac{x_{\tau n}^{\alpha n}}{n} + o(1), \quad (A.13)$$

where $q(\alpha, \tau, \xi)$ and $q_0(\alpha, \tau, \xi)$ are defined in (A.10) and (A.12).

*Remark* 6. In [5, formula (42)] the result similar to (A.13) is claimed (moreover, even with the equality $q(\alpha, \tau, \xi) = q_0(\alpha, \tau, \xi)$), but its proof is not sufficiently argumented (cf., for example, the derivation in [14, formula (B.21)] of a simpler upper bound for $x_{\tau n}^{\alpha n}$).

P r o o f. For fixed $n, w, j$ denote by $x_1 < x_2 < \ldots < x_j$ the roots of the polynomial $Q_j^{(n,w)}(x)$. Denote also
$$\rho_i = \frac{Q_j(i+1)}{Q_j(i)}.$$

Then
$$\log Q_j(i) = \log Q_j(0) + \sum_{k=0}^{i-1} \log \rho_k, \qquad Q_j(0) = \frac{(n-2j+1)}{(n-j+1)}\binom{n}{j}.$$

From [14, formula (i)] for $k < x_1$ we have
$$j \log\left[1 - \frac{1}{(x_1-k)^2}\right] < \log \frac{\rho_k}{\rho_{k-1}} = \sum_{l=1}^{j} \log\left[1 - \frac{1}{(x_l-k)^2}\right] < 0.$$

In particular, after simple estimates we get
$$1 < \frac{\rho_{k-1}}{\rho_k} < 1 + \frac{ej}{(x_1-k)^2}, \quad \text{if} \quad (x_1-k)^2 \geq j+2. \tag{A.14}$$

For $\{\rho_k\}$ the recurrent equation holds [14, formula (B.20)]
$$(w-k)(n-w-k)\rho_k \rho_{k-1} - b_k \rho_{k-1} + k^2 = 0,$$

where
$$b_k = w(n-w) - k(n-2k) - j(n+1-j).$$

Using representation $\rho_{k-1} = (1+\varepsilon_k)\rho_k$, where $\varepsilon_k$ is evaluated in (A.14), we have
$$(w-k)(n-w-k)(1+\varepsilon_k)\rho_k^2 - b_k(1+\varepsilon_k)\rho_k + k^2 = 0,$$

from which we get (since $\sqrt{a+b} \leq \sqrt{a} + \sqrt{b}$)
$$\rho_k \leq \frac{b_k + \sqrt{b_k^2 - 4k^2(w-k)(n-w-k)/(1+\varepsilon_k)}}{2(w-k)(n-w-k)} \leq$$
$$\leq \frac{b_k + \sqrt{b_k^2 - 4k^2(w-k)(n-w-k)}}{2(w-k)(n-w-k)} + \frac{k\sqrt{\varepsilon_k}}{\sqrt{(w-k)(n-w-k)}},$$

where
$$b_k^2 - 4k^2(w-k)(n-w-k) \leq$$
$$\leq [w(n-w) - j(n-j) - kn]^2 - 4k^2 j(n-j) + 2nkj.$$

*Remark* 7. The upper bound for $\rho_k$, perhaps, asymptotically (as $n \to \infty$) is exact, but it needs a more accurate investigation.

Denote
$$f_k = \sqrt{[w(n-w) - j(n-j) - kn]^2 - 4k^2 j(n-j)} \geq 0,$$
$$k \leq \frac{w(n-w) - j(n-j)}{n + 2\sqrt{j(n-j)}},$$

and
$$g_k = \frac{w(n-w) - j(n-j) - k(n-2k) + f_k}{2(w-k)(n-w-k)}.$$

Then we have
$$\rho_k \leq g_k + \frac{\sqrt{2nkj}}{2(w-k)(n-w-k)} + \frac{k\sqrt{\varepsilon_k}}{\sqrt{(w-k)(n-w-k)}},$$

and therefore
$$\log \rho_k \leq \log g_k + \delta_k,$$
$$\delta_k = \frac{\sqrt{2nkj} + 2k\sqrt{\varepsilon_k w(n-w)}}{w(n-w) - k(n-2k) - j(n-j)} \log e.$$

Now
$$\sum_{k=0}^{i-1} \log g_k = \sum_{k=0}^{i-1} \log \frac{w(n-w) - j(n-j) - k(n-2k) + f_k}{(w-k)(n-w-k)} - i.$$

We also have
$$\sum_{k=0}^{i-1} \log(w-k) \geq \int_0^i \log(w-x)\, dx =$$
$$= w \log w - (w-i)\log(w-i) - i \log e = wh_2\left(\frac{i}{w}\right) + i \log \frac{i}{e},$$

where the formula was used
$$bh_2\left(\frac{a}{b}\right) = b \log b - a \log a - (b-a)\log(b-a).$$

Then
$$\sum_{k=0}^{i-1} \log \frac{1}{(w-k)(n-w-k)} \leq -wh_2\left(\frac{i}{w}\right) - (n-w)h_2\left(\frac{i}{n-w}\right) - 2i \log \frac{i}{e}.$$

The function $T(x) = w(n-w) - j(n-j) - x(n-2x) + f_x$ is piecewise monotone, and it has not more than four intervals of monotonicity. Moreover, $\max_x T(x) = 2[w(n-w) - j(n-j)]$.
Therefore
$$\left| \sum_{k=0}^{i-1} \log T(k) - \int_0^i \log T(x)\, dx \right| \leq 4\log[w(n-w) - j(n-j)] + 4.$$

Hence

$$\sum_{k=0}^{i-1} \log g_k \leq \int_0^i \log T(x)\,dx - w h_2\left(\frac{i}{w}\right) - (n-w) h_2\left(\frac{i}{n-w}\right) - 2i \log \frac{i}{e} - i +$$
$$+ 4\log[w(n-w) - j(n-j)] + 4\,.$$

In particualr, with $w = \alpha n$, $i = \xi n$, $t = \tau n$ we have

$$\frac{1}{n}\sum_{k=0}^{i-1}\log g_k \leq q_0(\alpha, \tau, \xi) + \frac{8\log n}{n}\,.$$

It follows from (A.14) that $\varepsilon_k \leq ej(x_1 - k)^{-2}$. Therefore we also have

$$\frac{1}{n}\sum_{k=0}^{i-1}\delta_k \leq \sum_{k=0}^{i-1} \frac{\sqrt{2nkj} + 2k\sqrt{\varepsilon_k w(n-w)}}{[w(n-w) - j(n-j)]\sqrt{j(n-j)}} \log e \leq$$
$$\leq \frac{\sqrt{n}\,i^{3/2} + i^2\sqrt{ew(n-w)}/(x_1 - i)}{[w(n-w) - j(n-j)]\sqrt{n-j}} \log e \leq \frac{n^{5/2}}{(w-j)(n-w-j)(x_1 - i)}\,.$$

As a result, with $w = \alpha n$, $i = \xi n$, $j = \tau n$ we get

$$\frac{1}{n}\log Q_{\tau n}^{(n,\alpha n)}(\xi n) \leq q_0(\alpha, \tau, \xi) + \frac{2\sqrt{n}}{(\alpha - \tau)(1 - \alpha - \tau)(x_1 - \xi n)}\,. \tag{A.15}$$

Since the function $Q_{\tau n}^{(n,\alpha n)}(\xi n)$ monotone decreases for $\xi < x_1/n$, and the function $q_0(\alpha, \tau, \xi)$ is continuous and bounded on $\xi$, from (A.15) the inequality (A.13) follows. ▲

Replacing $2\xi$ by $\omega$, from (A.13) and (A.11) we get the Theorem 2 and (14). ▲